\documentclass[prl,reprint,superscriptaddress,floatfix,aps]{revtex4-2}
\usepackage{amsmath}
\usepackage{amsfonts}
\usepackage{graphicx}
\usepackage{physics}
\usepackage[usenames]{color}
\usepackage{xcolor}
\usepackage[colorlinks,linktocpage,bookmarks=false,citecolor=blue,linkcolor=red,urlcolor=blue]{hyperref}
\usepackage{hyperref}
\usepackage{cleveref}  

\def\be{\begin{equation}} \def\ee{\end{equation}}
\def\bea{\begin{eqnarray}} \def\eea{\end{eqnarray}}

\begin{document}

\title{Dissipation stabilizes Dicke Time Quasicrystals}
\author{Sk Anisur}
\affiliation{Harish-Chandra Research Institute, a CI of Homi Bhabha National Institute, Chhatnag Road, Jhunsi, Allahabad 211019}
\affiliation{Homi Bhabha National Institute, Training School Complex, Anushakti Nagar, Mumbai 400 094, India}

\author{Shuva Mondal}
\affiliation{Harish-Chandra Research Institute, a CI of Homi Bhabha National Institute, Chhatnag Road, Jhunsi, Allahabad 211019}
\affiliation{Homi Bhabha National Institute, Training School Complex, Anushakti Nagar, Mumbai 400 094, India}

\author{Tanaya Ray}
\affiliation{School of Physical Sciences, Indian Association for the Cultivation of Science, Kolkata 700032, India}

\author{Sayan Choudhury}
\email{sayanchoudhury@hri.res.in}
\affiliation{Harish-Chandra Research Institute, a CI of Homi Bhabha National Institute, Chhatnag Road, Jhunsi, Allahabad 211019}
\affiliation{Homi Bhabha National Institute, Training School Complex, Anushakti Nagar, Mumbai 400 094, India}
\date{\today}

\begin{abstract}
Quasi-periodic driving protocols provide a powerful route to realize novel non-equilibrium phases of matter beyond the Floquet paradigm. However, these protocols inevitably lead to infinite-temperature heat death in isolated systems, which poses a major challenge to their experimental realization. We demonstrate that dissipation can be harnessed to stabilize quasi-periodically driven systems, enabling the realization of robust non-equilibrium phases. Using the paradigmatic open Dicke model, we provide a blueprint for realizing a stable time quasicrystal (TQC) by Fibonacci driving. This TQC is characterized by a robust sub-harmonic quasi-periodic response that is dictated by, but qualitatively distinct from the external Fibonacci drive. By directly analyzing the time evolution in the thermodynamic limit, we establish the existence of TQC order in this system for a wide parameter regime. We trace the origin of the stability of the TQC to the attractor structure induced by dissipation. Strikingly, the TQC order persists in the deep quantum regime with as few as two qubits. We systematically study the dependence of the TQC lifetime, $\tau^{\ast}$, on the number of qubits and demonstrate that $\tau^{\ast}$ increases monotonically with the system size. Crucially, the TQC is not observed in the absence of dissipation. Our work establishes dissipation as a mechanism for stabilizing non-equilibrium phases of matter under quasi-periodic drive.
\end{abstract}

\maketitle
\textit{Introduction:}
The dynamics of driven many-body systems has been a topic of intense interest in recent years, especially in the context of periodic (Floquet) driving~\cite{zhan2024perspective, bukov2015universal, mori2023floquet, harper2020topology, rudner2020band, oka2019floquet, weitenberg2021tailoring}. On the one hand, Floquet protocols have been harnessed for various applications in quantum information science, such as state preparation~\cite{claeys2019floquet,ritter2025autonomous}, entanglement steering~\cite{gangopadhay2025counterdiabatic}, quantum sensing~\cite{iemini2024floquet,mishra2021driving,moon2024discrete,yousefjani2025discrete,bai2023floquet,shukla2025prethermal}, and quantum communication~\cite{engelhardt2024unified}. On the other hand, Floquet driving has been employed for the realization of dynamical phases of matter such as Floquet topological insulators~\cite{rudner2013anomalous,nathan2019anomalous,zhang2021anomalous,mukherjee2017experimental} and discrete time crystals in the closed ~\cite{Sacha_2018, Else2020DTC, RevModPhys.95.031001,PhysRevLett.116.250401,PhysRevB.94.085112, PhysRevLett.118.030401,PhysRevLett.117.090402,PhysRevLett.120.110603, PhysRevLett.120.140401,PhysRevB.97.184301, PhysRevLett.120.180602, PhysRevB.104.094308, munoz2022floquet, biswas2025floquet, PhysRevB.99.104303, PhysRevB.95.214307, atoms9020025, pizzi2021higher, PhysRevLett.127.090602, chandra2025integrable, sarkar2024time, PhysRevLett.130.120403, PhysRevB.109.104311, biswas2025discrete,PhysRevB.111.125159, rahaman2024time} and open dissipative systems \cite{PhysRevLett.120.040404,ray2024entanglement}. Crucially, these phases do not have any static analog, and they represent remarkable examples of intrinsically non-equilibrium order.

Intriguingly, some recent works have demonstrated that the catalog of non-equilibrium phases can be considerably extended by employing certain classes of aperiodic driving protocols~\cite{nandy2017aperiodically,mukherjee2020restoring,zhao2021random,liu2026prethermalization,e27060609,zhao2022anomalous,zhao2022suppression,mori2021rigorous,kumar2024prethermalization,dutta2025prethermalization,ghosh2025heating,ma2025stable,moon2025experimental}. A notable class of such aperiodic protocols is that of quasi-periodic driving, which has been employed for realizing prethermal phases of matter such as time quasi-crystals (TQCs)~\cite{PhysRevLett.120.070602, PhysRevX.10.021032, PhysRevB.100.134302,luo2025discrete,PhysRevX.15.011055, zhu2025observation} and temporal topological states. However, quasi-periodically driven systems generally heat up much faster than their Floquet counterparts due to the presence of multiple incommensurate frequencies in the drive. Consequently, these systems inevitably lead to an infinite-temperature heat death in isolated quantum systems~\cite{PhysRevLett.120.070602, PhysRevX.10.021032,PhysRevX.4.041048,pilatowsky2023complete,pilatowsky2025critically}. In this context, it is worth noting that dissipation can be harnessed to evade thermalization in Floquet systems~\cite{PhysRevLett.120.040404,ritter2025autonomous}. This raises a natural question: Is it possible to employ dissipation to stabilize non-equilibrium phases of matter in quasi-periodically driven systems?

In this Letter, we answer this question affirmatively by demonstrating the emergence of a robust TQC in the Fibonacci-driven open Dicke model (DM). These TQCs are characterized by a rigid quasi-periodic response that is significantly shifted from that of the drive. Notably, we establish the existence of a robust TQC both in the thermodynamic limit and in the deep quantum regime with as few as 2 qubits. Furthermore, we establish that the TQC lifetime increases monotonically with the system size. Crucially, in the absence of dissipation, the same drive protocol does not support a TQC, thereby establishing the fundamental role played by dissipation in stabilizing the TQC. Our results demonstrate that the dissipation provides a powerful route to realize non-equilibrium phases of matter in quasiperiodically driven systems.

\begin{figure*}
    \centering
    \includegraphics[width=1.0\linewidth]{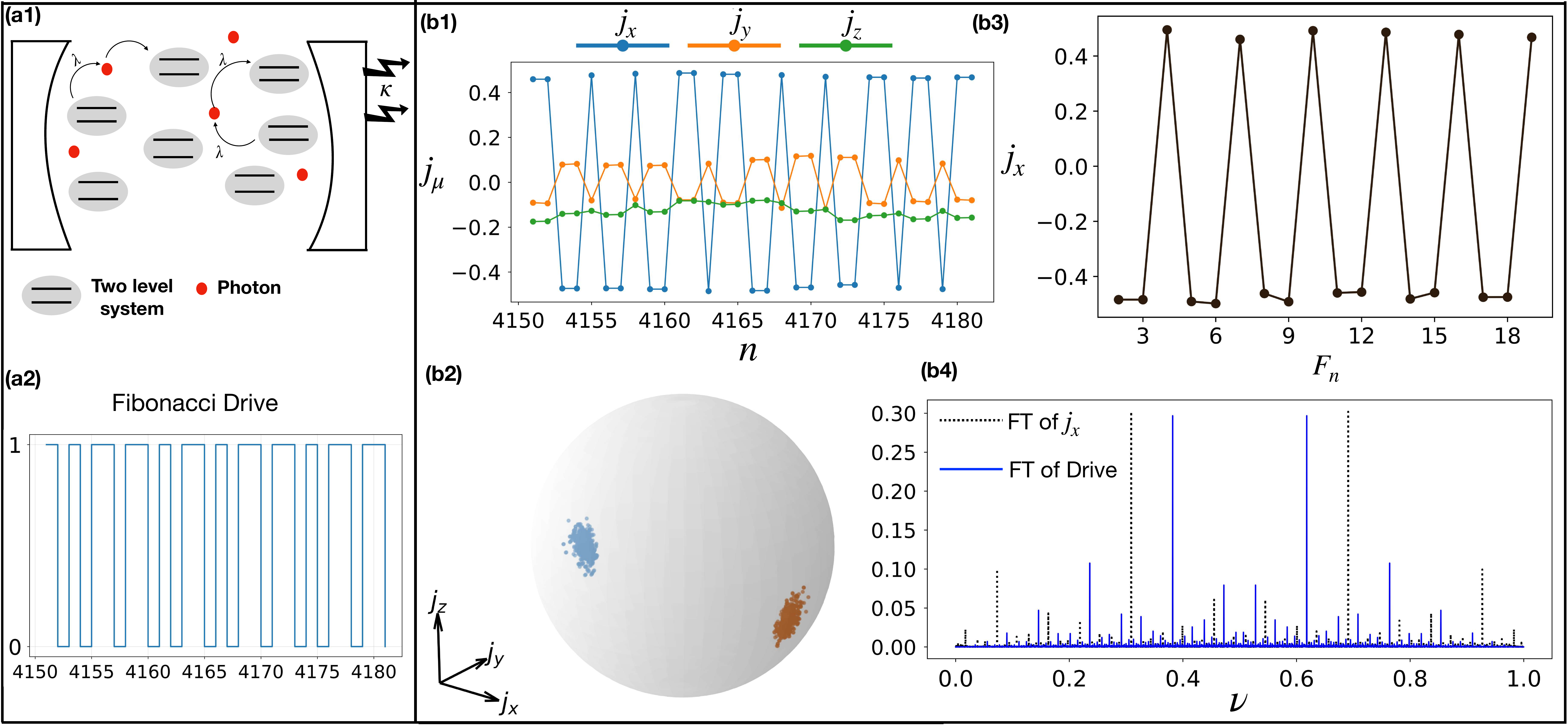}
    \caption{\textbf{Time quasicrystals in the open Dicke model:} (a1)-(a2): A schematic representation of the open DM describing the dissipative dynamics of $N$ two-level atoms coupled to a cavity, and the Fibonacci drive protocol that is studied in this work. (b1) The dynamics of the collective spin operators $j_{\alpha} = \langle J_{\alpha}\rangle/N$ at long times for $\epsilon=0$. The stable quasi-periodic oscillations exhibited by this system is a hallmark of the time quasi-crystal (TQC) phase. (b2) The phase-space trajectories projected on the pseudo-spin Bloch sphere for $\epsilon=0$. The robust quasi-periodic oscillations in the TQC phase is associated with transitions between the two stable islands. (b3) The TQC phase is characterized by period-three oscillations of $j_{x}$ in Fibonacci time. (b4) The Fourier transform the $j_{x}$ (black, dashed), and the drive (blue, solid) shows that the TQC response is shifted significantly from the drive.}
    \label{Fig:1}
\end{figure*}

\begin{figure}
    \centering
    \includegraphics[width=\linewidth]{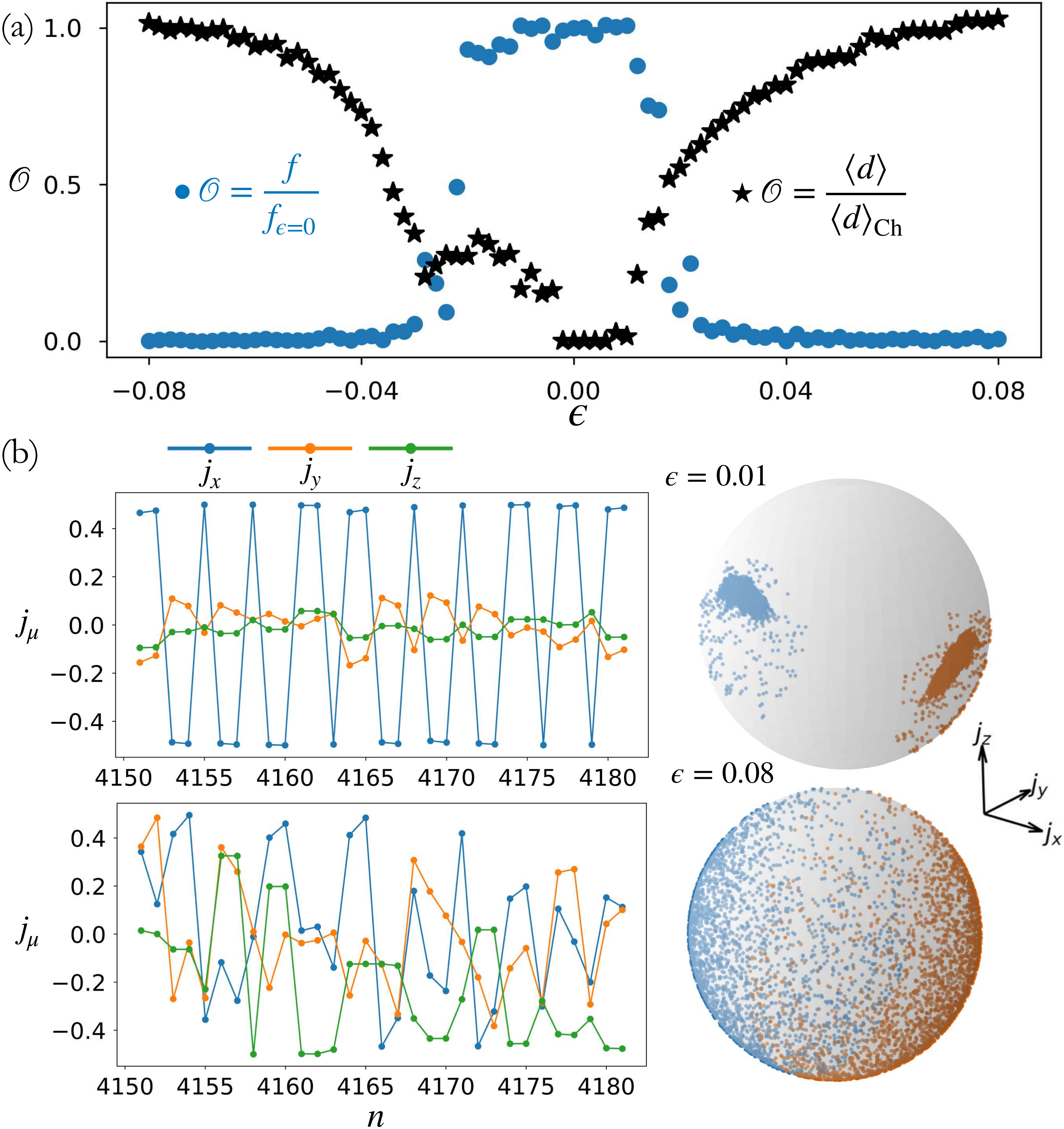}
    \caption{\textbf{Robustness of the TQC:} (a) The normalized TQC fraction $f / f_{\epsilon=0}$ (blue circles) and the normalized time-averaged decorrelator $\langle d \rangle/\langle d \rangle_{\mathrm{Ch}}$ (black stars) are employed to distinguish the TQC regime from the chaotic regime. The TQC phase is robust and it persists over a finite parameter regime, where, $\langle d \rangle_{\mathrm{Ch}}$ is the average of the decorrelator in the chaotic regime $(-0.08\le \epsilon \le -0.07)$ (b) The left panel illustrates the dynamics of the collective spin operators, $j_{\alpha} = \langle J_{\alpha}\rangle/N$ for representative values of $\epsilon$ in the TQC and the chaotic phase, and the right panel shows the corresponding phase-space trajectories on the Bloch sphere. The TQC and chaotic phases are characterized by regular and chaotic trajectories respectively, leading to consequent localization and delocalization on the Bloch sphere.}   
    \label{Fig:2}
\end{figure}

\textit{Model:}
We examine the dynamics of $N$ identical two-level atoms (qubits) in a single-mode cavity, described by the open DM \cite{PhysRevA.75.013804, PhysRevLett.120.040404, https://doi.org/10.1002/qute.201800043}:
\begin{equation}
\begin{aligned}
\frac{d \hat{\rho}_t}{dt}
&= \mathcal{L}(\lambda)\hat{\rho}_t = -i\bigl[\hat{H}(\lambda), \hat{\rho}_t\bigr]
+ \kappa \mathcal{D}[\hat{a}] \hat{\rho}_t, \\
\hat{H}(\lambda)
&= \omega \hat{a}^\dagger \hat{a}
+ \omega_0 \hat{J}_z
+ \frac{2\lambda}{\sqrt{N}}
(\hat{a} + \hat{a}^\dagger)\hat{J}_x .
\end{aligned}
\label{eq:Dicke}
\end{equation}
where $\mathcal{D}[\hat{a}]\hat{\rho} \equiv \hat{a}\hat{\rho}\hat{a}^\dagger
- \frac{1}{2}\left\{
\hat{a}^\dagger \hat{a}, \hat{\rho}
\right\}$, $\hat{a}(\hat{a}^{\dagger})$ is the annihilation (creation) operator of the photon field, and
$\hat{J}_\mu = \frac{1}{2}\sum_{j=1}^{N} \hat{\sigma}_j^\mu, (\mu = x,y,z)$ is the collective atomic pseudospin operator. Furthermore, $\omega_{0}, \omega, \kappa$ and $\lambda$ represent the atomic frequency, optical frequency, photon-loss rate and light-matter coupling strength respectively. This system exhibits a dissipative phase transition from a $\mathbb{Z}_2$-symmetric normal phase to a $\mathbb{Z}_2$-broken superradiant phase, when $\lambda = \lambda_c = \tfrac{1}{2}\sqrt{\frac{\omega_0}{\omega}  \left(\omega^2 + \kappa^2/4\right)}$~\cite{emary2003chaos}. The detuning between atomic and optical frequencies is parametrized by $\epsilon$ through $\omega = (1-\epsilon)\omega_{T}, \quad \omega_{0} = (1+\epsilon)\omega_{T}$ and we set $\omega_{T}=1$ throughout.

Intriguingly, a rich array of non-equilibrium phases emerge in this system under periodic driving. In particular, it is possible to realize time-translation-symmetry-breaking and subsequent discrete time crystal (DTC) order by periodically changing $\lambda$ from $0$ to $\lambda > \lambda_c $~\cite{PhysRevLett.120.040404}. These DTCs inherit their robustness from the presence of dissipation, and their stability can be further enhanced by rendering the dissipation non-Markovian~\cite{das2024discrete}. We now proceed to examine the phases that emerge in this system under quasi-periodic driving. 

To this end, we consider a quasi-periodic extension of the protocol employed to realize a DTC emerges under periodic driving ($\lambda = 1$ and $\kappa = 0.05$)~\cite{PhysRevLett.120.040404}:
\begin{equation}
\lambda_t =
\begin{cases}
\lambda \bigg((1+r_n)/2\bigg), &  (n-1)T \le t < (n-1/2)T, \\
0, & (n-1/2)T \le t < n T .
\end{cases}
\end{equation}
Here $\lambda > \lambda_c$ and $r_n$ is the $n-$th element of the binary Fibonacci sequence,  obtained from:
\begin{equation*}
    r_n = {\rm Sgn}\Bigg[ (\cos(2\pi b (n+1/2)) - \cos(b \pi))\Bigg],
\end{equation*}
where ${\rm Sgn}[x]$ denotes the signature of $x$, $b=1/\phi$ and $\phi = \frac{1+\sqrt{5}}{2}$ is the golden ratio. We now examine the emergence of a TQC in this model.

\textit{Semi-classical analysis:}
We first focus on the dynamics of the system directly in the thermodynamic limit, where the mean-field semiclassical equations are exact~\cite{bhaseen2012dynamics,carollo2021exactness}. Consequently, the time evolution of the quadratures $x = \frac{\langle \hat{a} + \hat{a}^\dagger \rangle}{\sqrt{2N\omega}}$, $ p = \frac{i\langle \hat{a}^\dagger - \hat{a} \rangle}{\sqrt{2N/\omega}}$ and the normalized collective spin components
$j_\mu = \langle \hat{J}_\mu \rangle / N$ $(\mu = x, y, z)$ with $\mathbf{j} \equiv (j_{x}, j_{y}, j_{z})$ are given by
\begin{equation}
\begin{aligned}
\frac{d\mathbf{j}}{dt}
&=
\left(
\omega_0 \mathbf{e}_z
+
2\lambda_t \sqrt{2\omega}\, x\, \mathbf{e}_x
\right)
\times \mathbf{j},
\\
\frac{dx}{dt}
&=
p-\frac{\kappa}{2}x,
\\
\frac{dp}{dt}
&=
-\omega^2 x
-\frac{\kappa}{2}p
-2\lambda_t \sqrt{2\omega}\, j_x .
\end{aligned}
\label{eq:evo}
\end{equation}
The steady states of the semi-classical equations are obtained by setting all time derivatives to zero. Intriguingly, the dissipative normal-superradiant phase transition is manifested as a dynamical phase transition. Below the critical light-matter coupling ($\lambda < \lambda_c$), the system has an unique attractor: $j_{x}=j_{y}=0$, $j_{z}= -1/2$, $x=p=0$. This attractor becomes unstable when $\lambda > \lambda_c$, and two new stable attractors emerge through a pitchfork bifurcation: $x = \mp x_s$ , $j_x = \pm j_{x,s}$ , $j_z \neq -\frac{1}{2}$ where $x_s = \mp \frac{\lambda}{\omega^2 + \kappa^2/4} \sqrt{2 \omega(1 - \frac{\lambda_c^4}
{\lambda^4})}$ , $p_s = \mp \frac{\kappa /2}{\omega^2 + \kappa^2/4} \sqrt{2 \omega(1 - \frac{\lambda_c^4}
{\lambda^4})}$ , $j_{x,s} = \pm \frac{1}{2}\sqrt{1 - \frac{\lambda_c^4}{\lambda^4}}$ , $j_{y,s} = 0$ , $j_{z,s} = -\frac{\lambda_c^2}{2\lambda^2}$. We note that a robust DTC can be realized by a Floquet protocol that periodically cycles this system between these two attractors. 

\begin{figure}
\centering
    \includegraphics[width=1.0\linewidth]{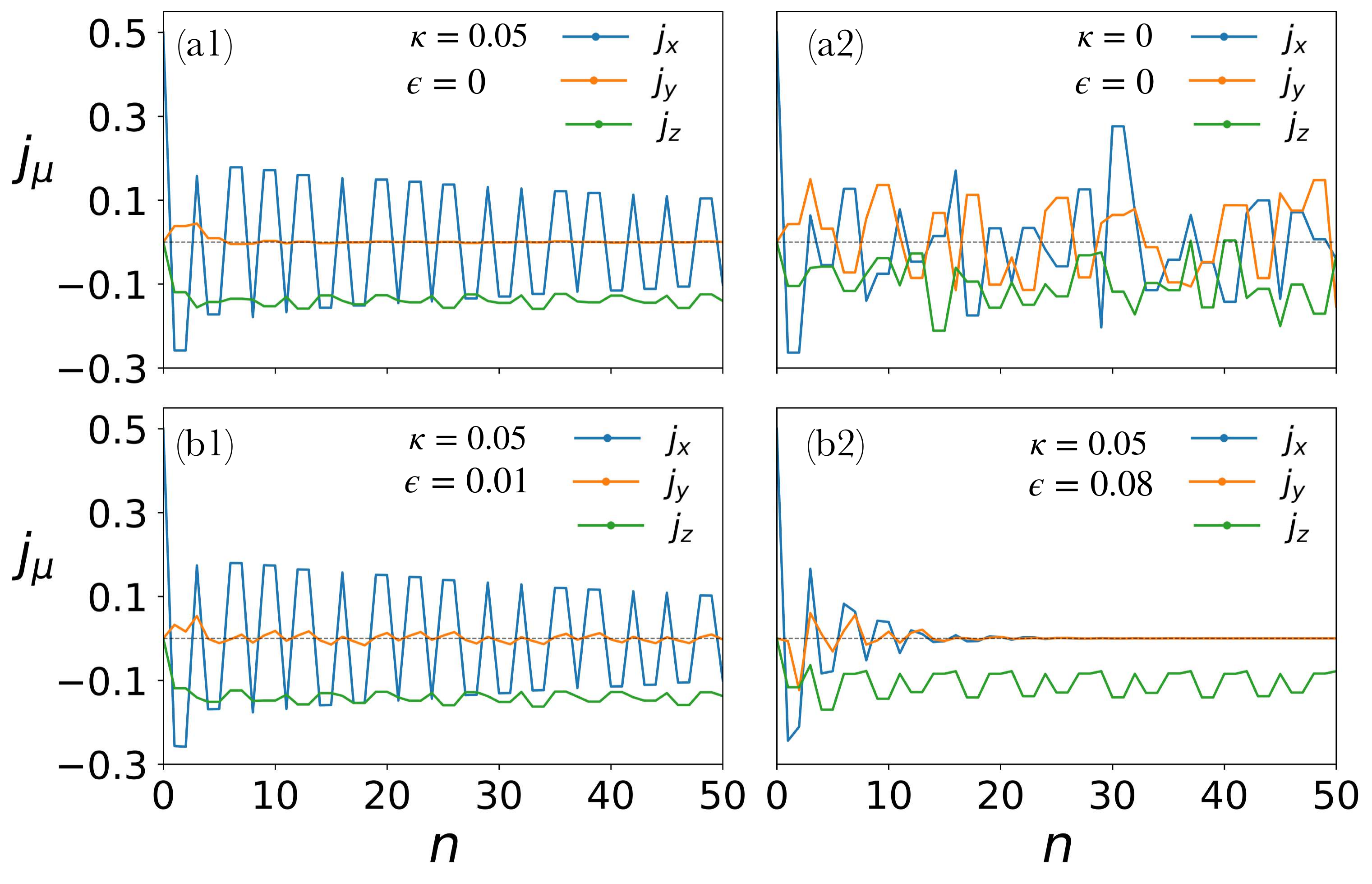}
    \caption{\textbf{TQCs in the deep quantum regime:} (a1)-(a2) The time-evolution of the Fibonacci driven two-qubit DM for $\epsilon = 0$ in the dissipative $[\kappa = 0.05]$ and closed $[\kappa = 0]$ regimes respectively. Persistent quasi-periodic oscillations are observed only in the presence of dissipation, thus demonstrating the key role played by dissipation in stabilizing the TQC. (b1)-(b2) Long-lived quasi-periodic oscillations persist in the small $\epsilon$ regime. However, these oscillations disappear when $\epsilon$ becomes large.}
    \label{Fig:3}
\end{figure}

We investigate the dynamics of the DM under the Fibonacci driving protocol, starting from a collective spin configuration in the vicinity of a superradiant phase attractor ($j_x^{(0)} = \tfrac{1}{2}\sqrt{1 -\lambda_c^4/\lambda^4}$, $j_y^{(0)} = 0$, and 
$j_z^{(0)} = -\tfrac{1}{2}\lambda_c^2/\lambda^2$). Figures~\ref{Fig:1}(b1)-(b2), show that the system exhibits stable quasi-periodic oscillations, and the system remains localized in stable islands around the two attractors. In this context, we note that a Fibonacci TQC has been investigated in a many-body localized spin chain, where a $\pi$-pulse was implemented at $t= \lfloor\phi m \rfloor$ ($m \in \mathbb{Z}^{+}$) to flip the magnetization quasi-periodically~\cite{PhysRevLett.120.070602}; this $\pi$-pulse played the role analogous to quasiperiodically modulated light-matter coupling $\lambda$ in our model. Consequently, the Dicke TQC  exhibits characteristic oscillations and a corresponding spectral response analogous to the TQC studied in Ref.~\cite{PhysRevLett.120.070602}. In particular,  $j_x$ exhibits period-3 oscillations in Fibonacci time, $F_n = (\phi^n - (-1)^{n} \phi^{-n})/\sqrt{5}$, where $\phi = (\sqrt{5}+1)/2$ is the golden ratio [Fig.~\ref{Fig:1}(b3)]. To see the origin of this Fibonacci-time periodicity, we note that the TQC phase is characterized by $F_{n-1}$ mod $2$ oscillations between the two attractors at $t=F_n$. As demonstrated in the Supplemental Material~\cite{suppmat}, the sequence $d_{n-1} = F_{n-1}$ mod $2$ exhibits a period-3 periodicity. This in turn leads to the characteristic Fibonacci-time oscillations. Crucially, the stroboscopic oscillations of the $j_x$ is quasi-periodic, but distinctly different from the drive pattern. This is reflected in the Fourier response of $j_{x}$, which is significantly shifted from that of the drive [Fig.~\ref{Fig:1}(b4)]. We emphasize, however, that in contrast to the prethermal nature of the many-body localized TQC~\cite{PhysRevLett.120.070602}, we establish the stability of the Dicke TQC directly in the thermodynamic limit. 

We distinguish the TQC and chaotic phases by employing two diagnostics - the quasi-crystalline fraction and the decorrelator. We start our analysis by computing the quasi-crystalline fraction~\cite{PhysRevX.15.011055},
\begin{equation}
f(\epsilon) = \frac{\left| S(\nu_0) \right|}
{\sum_{\nu_0 - \delta}^{\nu_0 + \delta} \left| S(\nu) \right|},
\label{eq:qc_fraction}
\end{equation}
where $\nu_0$ denotes the frequency of the dominant Fourier peak corresponding to $\epsilon=0$, and we set $\delta = \frac{1}{20}$ to avoid overlaps with other peaks. Our result shown in Fig.~\ref{Fig:2}(a) clearly demonstrates the existence of a robust parameter regime with a high crystalline fraction corresponding to a robust TQC.

We identify the chaotic regime by employing the decorrelator \cite{PhysRevB.108.024302,pizzi2021higher}:
\begin{equation}
\langle d \rangle = \frac{1}{t_f - t_i} \sum_{t=t_i}^{t_f} 
\left| \, |j_x(t)| - |j_x'(t)| \, \right| ,
\label{eq:decorrelator}
\end{equation}
where $j_x'(t)$ represents the time evolution originating from a slightly perturbed initial condition relative to $j_x(t)$. The perturbed initial state is chosen as $j_x'(0) = j_x(0) - 0.5 \times 10^{-3}$, $j_y'(0) = 0$, and $j_z'(0) = -\sqrt{\frac{1}{4} - |j_x'(0)|^2}$, ensuring the normalization of the spin vector. A high value of the decorrelator indicates sensitive dependence of initial conditions - a hallmark of chaos. Fig.~\ref{Fig:2}(a) shows the time-averaged decorrelator $\langle d \rangle$ over $5000$ driving periods. The TQC (chaotic) regime is  characterized by low (high) decorrelator value and a high (low) quasi-crystalline fraction. 

We emphasize that while the Fibonacci driving dictates the exact quasiperiodicity of the TQC oscillations, the stability of this phase originates from the dissipation induced attractor structure of the open DM. Crucially, in the TQC phase the trajectories in the Bloch sphere remain localized near the two $\mathbb{Z}_2$ broken fixed points, cycling between them in a regular quasi-periodic pattern [Figs.~\ref{Fig:1}(b2) and ~\ref{Fig:2}(b)]. In contrast, at $\epsilon = 0.08$, the dynamics of $j_{\mu}$ become irregular, and the trajectories spread across the Bloch sphere, which is the hallmark of chaotic behavior. These results establish the existence of the TQC directly in the thermodynamic limit. Furthermore, our analysis suggests that appropriately tailored driving protocols can harness this attractor structure to stabilize a broader class of non-equilibrium phases in the open DM. We note, however, that these results are obtained from the semi-classical equations of motion and are strictly valid in the thermodynamic limit. It is thus crucial to establish that the TQC survives in the presence of quantum fluctuations. We proceed to examine this next.

\begin{figure}
    \centering
    \includegraphics[width=1.0\linewidth]{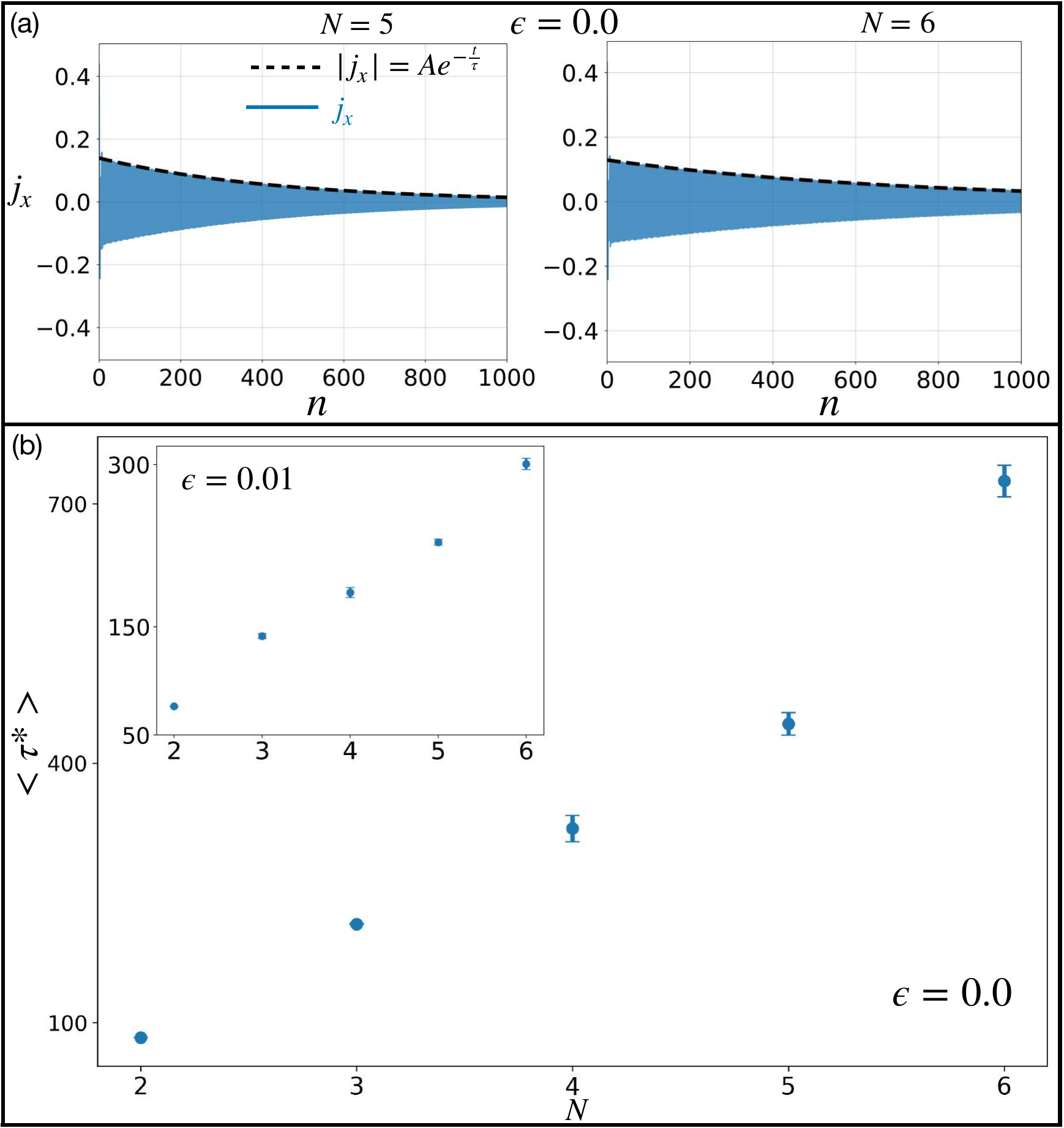}
    \caption{\textbf{TQC lifetime scaling:}  (a) The procedure to extract the TQC lifetime, $\tau$ by fitting $|j_x(t)| = A e^{-t/\tau}$ is illustrated for $N=5$ (18 photons) and $N=6$ (21 photons). (b) The dependence of the TQC lifetime on the number of qubits $N$ is shown. The errors bars have been obtained by obtaining $\tau$ for different truncation of photon numbers. Specifically, for $N=2,3,4$ qubits, we average over $13$ to $16$ photons, for $N = 5$ qubits over $15$ to $18$ photons, and for $N = 6$ qubits over $18$ to $21$ photons.  For $\epsilon = 0$, the lifetime increases monotonically with system size. The inset shows the corresponding behavior for $\epsilon = 0.01$, where the lifetime also grows with system size but at a slower rate compared to the $\epsilon = 0$ case. These results indicate that TQC order persists even in the deep quantum regime.}
    \label{Fig:4}
\end{figure}

\textit{Deep quantum regime:} In order to establish the stability of the TQC away from the thermodynamic limit, we examine the dynamics of the system in the few-qubit regime, which is often relevant for circuit-QED systems. We start our analysis by examining the two-qubit Dicke model and employ exact diagonalization to determine the dynamical evolution of this system. For these computations, the system is initially prepared in the \( |+\rangle^{\otimes N} \otimes |0\rangle \) state corresponding to both qubits polarized along the \(+x\) direction, $|0\rangle$ corresponds to the photon vacuum state. Figure~\ref{Fig:3} shows that even in this deep quantum regime, this system exhibits a long-lived TQC response in the TQC regime [Fig.~\ref{Fig:3}(a1)]. Dissipation is crucial for this stability and the TQC order does not persist in its absence [Fig.~\ref{Fig:3}(a2)]. Furthermore, we note that the TQC dynamics continues to persist in the small $\epsilon$ regime (away from $\epsilon=0$), thereby demonstrating the robustness of the TQC [Fig.~\ref{Fig:3}(b1)]. The TQC is destroyed in the large $\epsilon$ regime [Fig.~\ref{Fig:3}(b2)]. 

Having established the persistence of the TQC order even in the 2-qubit Dicke model, next we analyze the lifetime of the TQC as the number of qubits is increased. We obtain the TQC lifetime $\tau^{\ast}$ by fitting the time evolution of $j_x(t)$ to an exponentially decaying envelope of the form: $|j_x(t)| = A e^{-t/\tau^{\ast}}$, where $A$ is the initial amplitude. The fitting is performed after discarding the initial transient oscillations, so as to capture only the long-time decay behavior of the envelope. Furthermore, to ensure convergence in our calculations, we average over several photon truncations. Our results are shown in Fig.~\ref{Fig:4}. Remarkably, we find that $\tau^{\ast}$ increases monotonically with system size for both the detuning-free and detuned case, signaling the stability of the TQC phase against quantum fluctuations. We conclude that dissipation enables the realization of TQC order even in the deep quantum regime even in the presence of quasi-periodic driving.

\textit{Summary and outlook:} In this work, we have demonstrated that dissipation-induced attractors provide a mechanism to stabilize non-equilibrium phases of matter in quasi-periodically driven systems. Taking the paradigmatic example of the Fibonacci-driven open DM, we have proposed a route for realizing robust TQCs. The stability of these TQCs are endowed by the dissipation-induced attaractors, while the quasiperiodicity is dictated by the Fibonacci drive. Notably, despite the presence of multiple incommensurate frequencies in the drive, the TQC phase is stable over a wide parameter regime in the thermodynamic limit. Furthermore, we have established that long-lived  TQC dynamics can be observed even in the deep quantum regime with as few as 2 qubits. Crucially, this TQC order persists only in the presence of dissipation. Finally, we have analyzed the dependence of the TQC lifetime, $\tau^{\ast}$ on the number of qubits, and demonstrated that $\tau^{\ast}$ increases monotonically with $N$. Our results pave the path towards employing dissipation for realizing stable dynamical phases in quasi-periodically driven systems, taking an important step beyond previously studied protocols in isolated systems, where such phases were prethermal. 

There are several promising venues for future work. A natural first step would be to examine the array of stable non-equilibrium phases that emerge in other kinds of open Dicke models~\cite{das2023periodically,das2026detailed,yegovtsev2026robust,jachinowski2026spin} and Dicke dimers in the presence of structured aperiodic driving. Furthermore, it would be interesting to investigate the stability of the TQC phase when the dissipation becomes non-Markovian~\cite{lundgren2020nature,muller2025quantum}. Finally, we note that a recent work has provided a pathway to harness dissipation to stabilize DTCs in a non-collective many-body system~\cite{raman2026dissipative}. Generalizing such a strategy to aperiodically driven systems would be an intriguing direction for future research.\\

\emph{Acknowledgements:} SC acknowledges funding under the Government of India’s National Quantum Mission grant numbered DST/FFT/NQM/QSM/2024/3.

\bibliography{ref}

\onecolumngrid

\pagebreak

\onecolumngrid

\begin{center}
{\Large Supplemental Material for: `Dissipation-Stabilized Time Quasicrystals in the Open Dicke Model'}
\end{center}

In this Supplemental Material, we provide the derivations of Eq.~(3) in the main text and the procedure to determine the lifetime of the TQCs.

\section{Derivation of the Semiclassical Equations of Motion}
\label{app:derivation}

Here we derive the semiclassical equations of motion presented in Eq.~\ref{eq:evo} of the main text. For an open quantum system, the Heisenberg equation of motion for a generic observable $\hat{O}$ is given by \cite{breuer2002theory},

\begin{equation}
\frac{d\langle \hat{O} \rangle}{dt} = \langle \mathcal{L}_t^\dagger \hat{O} \rangle = \Biggl\langle i\bigl[\hat{H}(t), \hat{O}\bigr] + \sum_j \Biggl( \hat{L}_j^\dagger(t)\, \hat{O}\, \hat{L}_j(t)  - \frac{1}{2} \left\{ \hat{L}_j^\dagger(t)\hat{L}_j(t), \hat{O} \right\} \Biggr) \Biggr\rangle. 
\end{equation}
This leads to the following time-evolution for $\hat{J}_x$:
\begin{equation*}
\frac{d\langle \hat{J}_{x} \rangle}{dt} = \Biggl\langle i\bigl[\hat{H}, \hat{J}_{x}\bigr] + \kappa \Biggl(\hat{a}^\dagger\, \hat{J}_{x}\, \hat{a} \ - \frac{1}{2} \left\{
\hat{a}^\dagger \hat{a}, \hat{J}_{x} \right\} \Biggr) \Biggr\rangle 
\end{equation*}

Proceeding further, we note that
\begin{equation*}
\left[ \hat{H}, \hat{J}_{x} \right] = \omega_{0} [\hat{J}_{z} , \hat{J}_{x}] + \frac{2\lambda}{\sqrt N} (\hat{a} + \hat{a}^\dagger) [\hat{J}_{x} , \hat{J}_{x}] = i \omega_{0} \hat{J}_{y}\,  
\end{equation*}
and 
\begin{equation*}
\begin{aligned}
\hat{a}^\dagger\, \hat{J}_{x}\, \hat{a}
- \frac{1}{2}\left\{\hat{a}^\dagger\hat{a}, \hat{J}_{x}\right\} = 0
\end{aligned}
\end{equation*}
These considerations lead to the following time evolution for $j_x = \langle \hat{J}_x/N \rangle$
\begin{equation}
    \frac{d {j}_{x}}{dt} = - \omega_{0} {j}_{y} 
\end{equation}
Similarly, we find that
\begin{equation*}
\frac{d\langle \hat{J}_{y} \rangle}{dt}= i \langle [\hat{H} , \hat{J}_{y}] \rangle,
\end{equation*}
where
\[
\begin{aligned}
\left[ \hat{H}, \hat{J}_{y} \right]
&= \omega_{0} [\hat{J}_{z} , \hat{J}_{y}] + \frac{2\lambda}{\sqrt N} (\hat{a} + \hat{a}^\dagger) [\hat{J}_{x} , \hat{J}_{y}] \\
&= - i \omega_{0} \hat{J}_{x}\,  +  \frac{2i\lambda}{\sqrt N} (\hat{a} + \hat{a}^\dagger)  \hat{J}_{z}.
\end{aligned}
\]
Consequently, we obtain
\begin{equation}
    \frac{d {j}_{y}}{dt} =  \omega_{0} {j}_{x} - 2\lambda \sqrt{2 \omega} x {j}_{z}.
\end{equation}
Finally,  we note that the time-evolution of $\hat{J}_{z}$ is governed by
\begin{equation*}
\frac{d\langle \hat{J}_{z} \rangle}{dt} = i \langle [\hat{H} , \hat{J}_{z}] \rangle = - \frac{2i\lambda}{\sqrt N} (\hat{a} + \hat{a}^\dagger)  \hat{J}_{y} 
\end{equation*}
We conclude that
\begin{equation}
    \frac{d {j}_{z}}{dt} = 2\lambda \sqrt{2 \omega}  x {j}_{y}.
\end{equation}
We now proceed to obtain the time-evolution of $x$ and $p$. To do this, we first note:
\begin{equation*}
\frac{d\langle \hat{a} \rangle}{dt} = i \Biggl\langle \bigl[\hat{H}, \hat{a} \bigr]
+ \kappa \Biggl(
\hat{a}^\dagger\, \hat{a}\, \hat{a} - \frac{1}{2}
\left\{
\hat{a}^\dagger \hat{a} , \hat{a}
\right\}
\Biggr)
\Biggr\rangle ,
\end{equation*}
where
\begin{equation*}
\left[ \hat{H}, \hat{a} \right] = - (\omega \hat{a} + \frac{2\lambda}{\sqrt N} \hat{J}_{x}) 
\end{equation*}
From these consideration, we conclude
\begin{equation}
    \frac{d\langle \hat{a} \rangle}{dt} = -i \omega \langle \hat{a} \rangle - \frac{2i\lambda}{\sqrt{N}} \langle \hat{J}_{x} \rangle - \frac{\kappa}{2}  \langle \hat{a} \rangle,
\end{equation}
and
\begin{equation}
    \frac{d\langle \hat{a}^\dagger \rangle}{dt} = i \omega \langle \hat{a}^\dagger \rangle + \frac{2i\lambda}{\sqrt{N}} \langle \hat{J}_{x} \rangle - \frac{\kappa}{2}  \langle \hat{a}^\dagger \rangle.
\end{equation}
These results in turn leads to:
\begin{equation}
    \frac{d x}{dt} = p - \frac{\kappa}{2} x
\end{equation}
\begin{equation}
    \frac{d p}{dt} = - \omega^2 x - 2 \sqrt{2 \omega} \lambda j_{x} - \frac{\kappa}{2} p
\end{equation}
The coupled nonlinear differential equations derived here (and shown in Eq.~\ref{eq:evo} in the main text are then numerically integrated using a fourth-order Runge–Kutta method to obtain the dynamics of ${\bf j} = \{j_x,j_y,j_z\}$. 

\section{Period-3 Recurrence in fibonacci time}
In the main text, we had noted that the sequence $d_{n-1} = F_{n-1}$ mod $2$ exhibits period-3 oscillations. To see this, we note that this $d_{n-1}$ takes the form:
\begin{equation}
d_{3n-2} = d_{3n-1} = 1, \,\, d_{3n}=0,
\end{equation}
where the pattern \{1,1,0\} repeats exactly every three steps. This result follows from the fact that, modulo 2, $F_{n+1} = F_n + F_{n-1}$ is fully determined by two consecutive values; since there are only four such possible pairs, the sequence must cycle between them. Starting from $(F_1, F_2)$ mod $2$ = $(1,1)$, we obtain the following sequence: $(1,1) \rightarrow (1,0) \rightarrow (0,1) \rightarrow (1,1)$  The pair $(1,1)$ thus recurs exactly after 3 steps, confirming that the period is 3. Consequently, in the TQC phase the system exhibits period-$3$ oscillations in Fibonacci time, as shown in Fig.~1(b3).

\end{document}